\documentclass[namedreferences]{SolarPhysics}      
\usepackage[optionalrh]{spr-sola-addons}

\usepackage{graphicx}                
\usepackage{multirow}
\usepackage{multicol}
\usepackage{fancyvrb}
\usepackage{lineno}
\usepackage{rotate}

\usepackage[usenames]{color}
\definecolor{myDarkGreen}{rgb}{0,0.64,0}
\definecolor{myDarkRed}{rgb}{0.64,0,0}
\definecolor{myGrey}{rgb}{0.64,0.64,0.64}
\definecolor{myBlack}{rgb}{0,0,0}
\usepackage{url}             

\def\GPR{GPR}
\def\GPRu{GPR(unreliable)}

\def\GPRl{GPR(linked)}
\def\GPRlp{GPR(linked$'$)}
\def\GPRlr{GPR(linked, reliable)}
\def\GPRt{GPR(training)}
\def\RGO{Royal Greenwich Observatory}
\def\textdeg{$^\circ$}

\def\@thedoi{10.1007/\textbullet\textbullet\textbullet\textbullet\textbullet-\textbullet\textbullet\textbullet-\textbullet\textbullet\textbullet-\textbullet\textbullet\textbullet\textbullet-\textbullet}

\begin{document}


\begin{article}

\begin{opening}

\title{Increasing Lifetime of Recurrent Sunspot Groups within the Greenwich
Photoheliographic Results}

\author{R.~\surname{Henwood}$^{1,2}$\sep
        S.C.~\surname{Chapman}$^{1}$\sep
        D.M.~\surname{Willis}$^{1,3}$
    }

\runningauthor{Henwood \emph{et al.}}
\runningtitle{Lifetime of Recurrent Sunspot Groups}

\institute{$^{1}$ 
    Centre for Fusion, Space and Astrophysics, University of Warwick, UK\\
    $^{2}$ 
    UK Solar System Data Centre, Rutherford Appleton Laboratory,
    Chilton, Didcot, Oxfordshire OX11 0QX, UK
        email: \url{Richard.Henwood@stfc.ac.uk}\\
    $^{3}$ 
    Space Science and Technology Department, Rutherford Appleton Laboratory,
    Chilton, Didcot, Oxfordshire OX11 0QX, UK
    }

\begin{abstract}

Long-lived ($>$20 days) sunspot groups extracted from the Greenwich
Photoheliographic Results (\GPR) are examined for evidence of decadal change.
The problem of identifying sunspot groups which are observed on consecutive
solar rotations (recurrent sunspot groups) is tackled by first constructing
manually an example dataset of recurrent sunspot groups and then using machine
learning to generalise this subset to the whole \GPR. The resulting dataset of
recurrent sunspot groups is verified against previous work by A. Maunder and
other \RGO\ (RGO) compilers. Recurrent groups are found to exhibit a slightly
larger value for the Gnevyshev-Waldmeier Relationship than the value found by
Petrovay and van Driel-Gesztelyi (\emph{Solar Phys.} \textbf{51}, 25, 1997), who used
recurrence data from the  Debrecen Photoheliographic Results. Evidence for
sunspot group lifetime change over the previous century is observed within
recurrent groups. A lifetime increase of 1.4 between 1915 and 1940 is found,
which closely agrees with results from Blanter {\it{et al.}} (\emph{Solar Phys.}
\textbf{237}, 329, 2006).  Furthermore,
this increase is found to exist over a longer period (1915 to 1950) than
previously thought and provisional evidence is found for a decline between
1950 and 1965. Possible applications of machine-learning procedures to the
analysis of historical sunspot observations, the determination of the magnetic
topology of the solar corona and the incidence of severe space-weather events
are outlined briefly.

\end{abstract}
\end{opening}

\section{Introduction}
\label{sec:intro}

The influence of the Sun on the Earth has attracted renewed attention in the
context of climate change \cite{svensmark1997}. Sunspots are a good measure of
solar activity and have been observed systematically for hundreds of years.
The Royal Greenwich Observatory (RGO) began an effort to record the position
and size of sunspot groups in 1874 and maintained this programme of solar
observations until the end of 1976. The resulting dataset is unrivalled in its
longevity and homogeneity. 

Several attempts to model the total quantity of solar radiation arriving at
the Earth (the total solar irradiance) have been undertaken using various
indices \cite{lean1995,fligge1997,Balmaceda2007}. Some of these attempts have
relied on the measurements of sunspot number, since this index extends back
for a few centuries. Modern high resolution imaging and measurement of sunspot
properties are of limited use because the characteristic times of solar
change, on top of the 22-year solar cycle, are expected to take place on
centennial time scales \cite{blanter2006}.

Studies of the temporal properties of sunspot groups (lifetime, maximum size,
heliographic position, etc) are hampered by two factors:  short-lived sunspot
groups may be missed due to nightfall \cite{solanki2003} and the rotation of
the Sun carries groups out of view from an Earth-bound observer. In addition,
the effects of fore-shortening and limb darkening will hamper reliable
observation away from the central meridian \cite{Pierce1977}.

\begin{figure}
\begin{center}
\scalebox{1.0}{\input{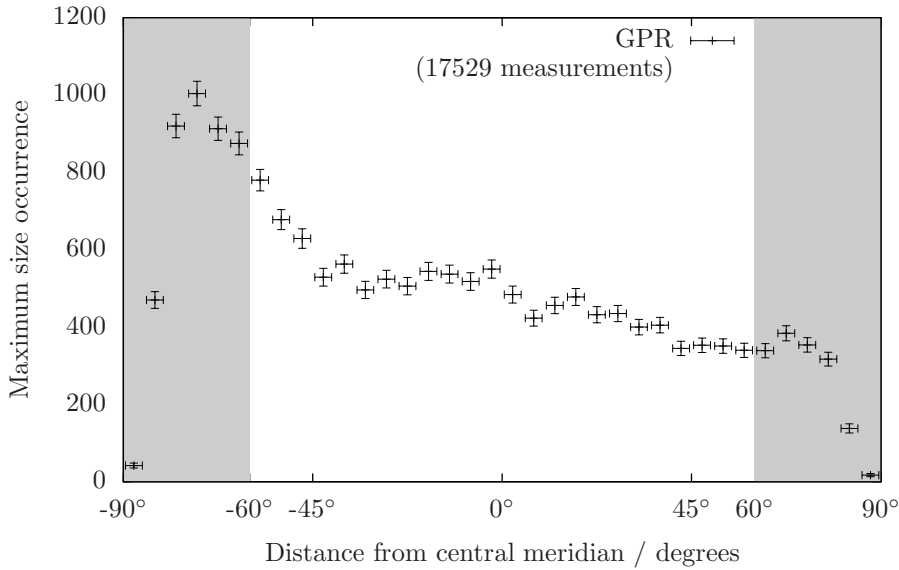}}

\caption{The longitudinal distance from the central meridian of the Sun at
which a sunspot group achieves maximum wholespot size. Error limits are the
width of the bins (5$^\circ$) in the x-direction and the square-root of the
number of elements in the \emph{y}-direction. \GPR\ data are filtered to remove
sunspot groups with only one observation, since these do not make meaningful
maximum-size contributions. The resulting \GPR\ data contain a maximum at
around -75$^\circ$.  This is attributed to long-lived spots which are growing
as they reach the west limb of the Sun or declining as they appear over the
east limb of the Sun. The $\chi^2$ value for this distribution
gives a high statistical significance ($>$99\%) to the departure from
uniformity. The declining measurements within the grey regions indicate that
reliable observations towards the solar limbs are difficult.}

\label{fig:unfilteredMaxSizeCMD}
\end{center}
\end{figure}

In order to quantify the limits of reliable observation of sunspot groups
within the \GPR, the longitude distribution of the apparent maximum size is
presented in Figure \ref{fig:unfilteredMaxSizeCMD}. From this illustration one
can conservatively conclude that observations at distances greater than
60$^\circ$ from the central meridian are difficult. This result is consistent
with theoretical findings \cite{kopecky1985}.  

Sunspot groups that are recorded at any stage with a central meridian
distance $<$-60$^{\circ}$ or $>$+60$^{\circ}$ are classified as `unreliably
observed'. This subset is named \GPRu\ herein.

The considerable asymmetry of the distribution in Figure
\ref{fig:unfilteredMaxSizeCMD} may be explained by a combination of sunspot
group decay rate and recurrence. This matter is briefly treated by
\inlinecite{henwood2008}. However, a detailed investigation is outside the
intended scope of this paper.

\inlinecite{blanter2006}, performed a non-linear study of short-term correlation
properties of solar activity in order to reveal long-lifetime variations.
This method was applied to the \GPR\ and an increase of lifetime by a factor
of 1.4 was observed from 1915 to 1940. A dataset of sunspot group lifetimes
that are not truncated by solar rotation would allow direct measurement of
lifetime and hence verify the observation of \citeauthor{blanter2006}.

In this study a training dataset of recurrent sunspot groups is constructed by
hand from longitude-time plots of \GPR\ data. This subset of \GPR\ data is
balanced to ensure an equal number of linked and unlinked examples, and
presented as training input to a number of feed-forward neural network
architectures. Each network is trained using 10-fold cross validation and the
network that displays the lowest over or under fitting is selected.  The
chosen network is then trained and applied to the entire \GPR\ dataset.
Finally, a simple filter is applied to the linked dataset and validation is
performed.

\section{Method}
\label{sec:method}

A meticulous observer can follow a long-lived sunspot group for successive
days until it disappears from view over the west limb of the Sun. By taking
note of the latitude of the last reliable observation, and with knowledge of
the solar rotation period, the observer could wait to see if a sunspot group
appeared over the east limb at roughly the same latitude and at the
appropriate time.  If such a prediction is fulfilled, one may conclude that
the same sunspot group has been observed on consecutive rotations and that it
should not be recorded as two separate sunspot groups. Ideally, a recurrent
sunspot group should be identified as such, possibly by using the same sunspot
group number for such recurrent observations. Within the digital version of
the \GPR, no recurrent information is recorded. Sunspot observations are
grouped together and allocated a unique Greenwich group number if the group is
observed on consecutive days. 

\inlinecite{becker1955} studied \emph{Sonnenfleckenherde}, which may be
translated as `focus of sunspots'. The method employed to identify recurrence
in that study was a ``statistical method''. \emph{Sonnenfleckenherde} were
defined as ``an area on the Sun in which, during a longer period of time, spots
appear or are being built''. Size, heliographic latitude and longitude are all
taken into consideration during the search for recurrent groups. This method
was applied to \GPR\ data from 1879 to 1941 and 46 \emph{Sonnenfleckenherde}
were catalogued.

\begin{figure}
\begin{center}
\scalebox{1.0}{\input{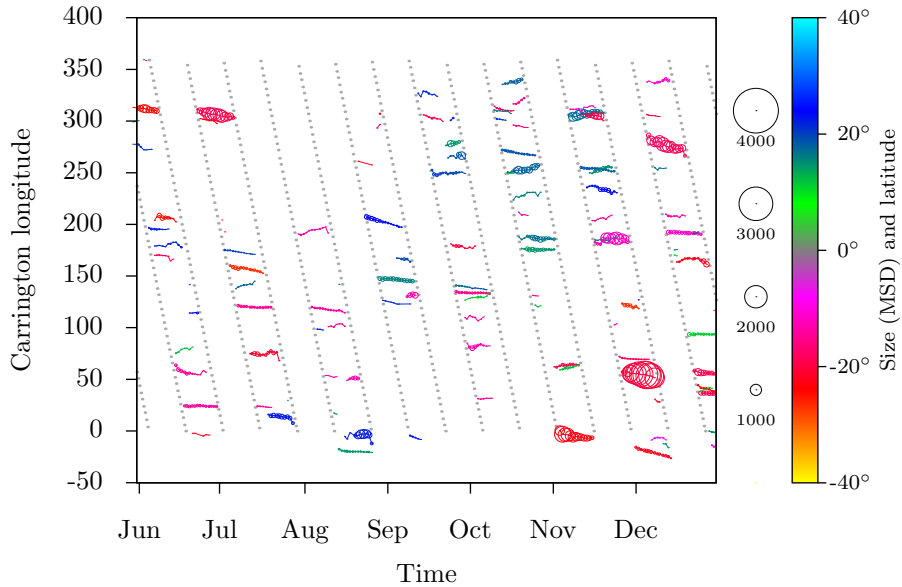}}

\caption{A longitude-time diagram of \GPRu\ groups for the last six months of
1935. Time is in the \emph{x}-direction, longitude in the \emph{y}-direction. \GPRu\ groups
are displayed as a connected series of coloured circles.  Colour corresponds to
latitude (right-hand colour bar); radius indicates size, corrected for
foreshortening and measured in millionths of the solar hemisphere. Example
radii are provided for sunspot sizes of 1000, 2000, 3000 and 4000 millionths
of the solar hemisphere (left of colour bar). Slanting grey dots mark the
Carrington longitude of the east and west limb of the Sun.}

\label{fig:lattime}
\end{center}
\end{figure}

Subsequently, \inlinecite{castenmiller1986} introduced the term
~\emph{sunspot nest} to describe the same phenomenon on the Sun. They defined
nests as evolving within one month, lasting for 6 to 15 rotations and not
expanding in latitude or longitude.  In their study, the primary means of
investigating sunspot nests was by visualising the \GPR\ in heliographic
longitude-time diagrams (Figure \ref{fig:lattime}) and recording recurrent
spots manually.  Practically, \inlinecite{castenmiller1986} required the
following data for their analysis: Carrington longitude, latitude and number
of ``visible days'' a group was observed during one solar rotation.  This method
was applied to the period August 1959 -- December 1964 of \GPR\ data, and 41
probable sunspot nests were found.

Within this paper a less stringent concept of a sunspot nest is employed,
namely a sunspot \emph{nestlet}. A \emph{nestlet} is defined as two or more
`unreliably' observed sunspot groups which are linked together because they
are likely to be the same group but have different group numbers in the \GPR.
Hence a \emph{nestlet} physically corresponds to a white-light sunspot group
observed on consecutive solar rotations. It is possible \emph{Nestlets} are
found preferentially in ``active longitudes'' or ``hot spots'' zones, but such
a study is outside the intended scope of this paper.

The method employed to find all the nestlets within the \GPR\ is as follows:

\begin{enumerate}
\item Construct a longitude-time diagram for a chosen solar cycle.
\item By inspection, create a list of recurrent sunspot groups.
\item Use this list to train a neural network.
\item Apply the trained neural network to the whole \GPR\ dataset.
\item Post-process to remove outliers.
\end{enumerate}

\subsection{Solar Differential Rotation}
\label{movement}

Sunspot groups move on the photosphere with different speeds depending on
latitude because of differential solar rotation
\cite{schroeter1985,thompson1996}. According to the annals published by the
Royal Greenwich Observatory, ``it should be noted that longitudes are based on
the ephemeris given in the Astronomical Ephemeris, assuming a solar rotation
period constant at all latitudes'' \cite{rgo1980}.  Hence, after an interval of one rotation,
recurring groups would be expected to show a drift in longitude due to
differential solar rotation.  Using the analysis of solar rotation from \GPR\
data, performed by \inlinecite{balthasar1986}, one should expect a group at an
extreme latitude of 35\textdeg\ to exhibit a backwards drift of at most
1\textdeg day$^{-1}$ with respect to the equator.  This would correspond to a drift
of not more than 18\textdeg\ during the whole unreliable passage (18 days).
The figure of 18 days is derived using a synodic solar rotation period of 27
days and an unseen passage of 180\textdeg\ (the far-side of the Sun) + the
unreliable regions of the near side of the Sun (2 $\times$ 30\textdeg).  This
gives: $\frac{2}{3}\times 27$ days = 18 days.

\begin{figure}
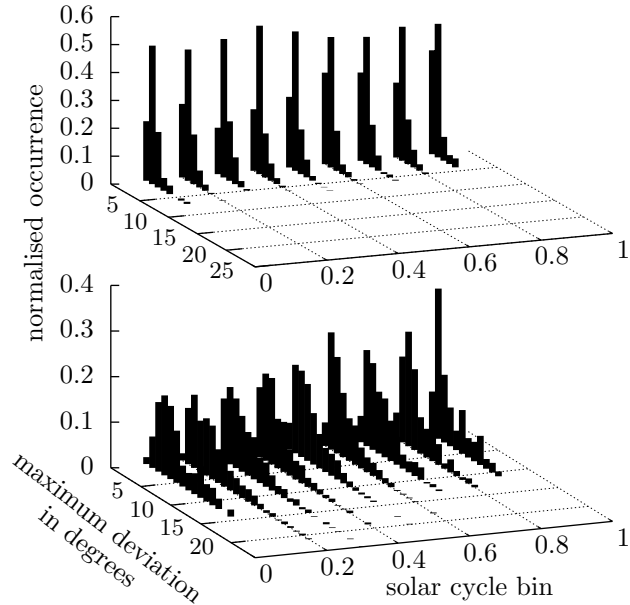

\begin{center}
\scalebox{1.0}{\input{./latitude3d.txt}}
\vskip -2.5cm
\scalebox{1.0}{\input{./longitude3d.txt}}
\vskip -0.5cm

\caption{Extreme deviation is measured for all sunspot groups within the \GPR\
that are observed to exist for more than ten days. The extreme deviation is
measured as the difference between the absolute maximum and minimum recorded
heliographic positions (``maximum deviation'', \emph{x}-direction). Deviation
counts are binned into solar cycle bins. Each solar cycle is divided into nine
bins (``solar cycle bin'', \emph{y}-direction). Deviation counts are
normalised (``normalised occurrence'', \emph{z}-direction). The extreme
latitude deviation (top) illustrates that groups commonly move a few degrees
but rarely more than five degrees.  Extreme longitude (bottom) deviations
indicate a greater spread but with groups rarely moving more than 15$^\circ$.}

\label{fig:spotMovement}
\end{center}
\end{figure}

Figure \ref{fig:spotMovement} demonstrates that sunspot groups rarely move more than
5$^\circ$ in latitude and 15$^\circ$ in longitude over the duration of their
observed life. Such movement appears to be independent of solar cycle
development. Hence the worst case of 18\textdeg\ longitude drift due to 
differential rotation is uncommon.

\subsection{Presenting \GPR\ Data to the Neural Network}
\label{sec:present}

The purpose of the proceeding approximate analysis of sunspot movement is to provide
boundary conditions when converting \GPR(unreliable)\ data into a form
suitable for the neural network. This process consists of two stages. First,
selecting groups that are possibly linked. Second, reducing each possible
linkage into a suitable network input.

For the first stage, the group movement metrics calculated above provide
approximate boundary conditions when considering if any two sunspot groups
could be linked. It is asserted here that for such a link to exist from an
initial group, the linked group (post group) must appear within latitude and
longitude bounds of $\pm$15\textdeg\ and $\pm$50\textdeg\ respectively. These
bounds are at the extreme end of observed group movement against an average
solar rotation (Section \ref{movement}).

In addition, a post group must appear in the future (with respect to the
initial group) and within a time window that starts when the first longitude
bound appears at the east limb of the Sun, and ends when the last longitude
bound appears at the east limb of the Sun. These bounds are calculated using
an ephemeris \cite{1991meeus}. A deliberately generous window is chosen
because decision making regarding recurrence is performed by the neural
network, not during pre-processing of the data.

Every group which is ever observed near the unreliable west limb of the Sun
has this first stage applied to it and a set of potential linked candidates
are created. Once a pair of potential linked candidates have been identified,
they are encoded for presentation to the neural network. In the first step,
the two groups which make up the linking candidates are classified as an
``initial group'' and a ``post group''. The ``initial group'' is the one that
is observed first in time and disappears over the west limb of the Sun. The
``post group'' is the one that is observed later in time, and emerges over the
east limb of the Sun.

\begin{figure}
\begin{center}
\scalebox{1.0}{\input{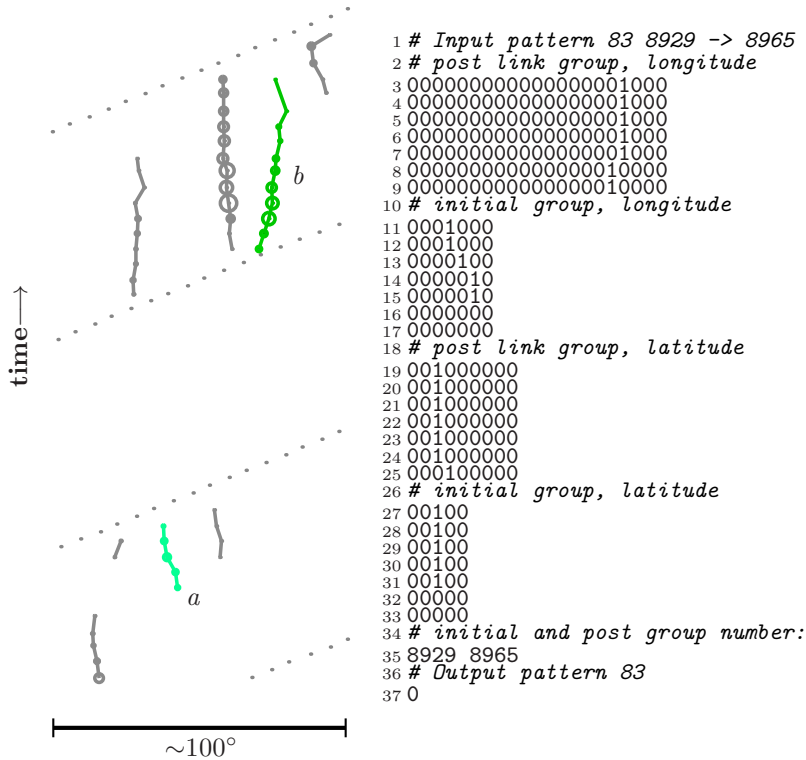}}

\caption{Schematic example of encoding a link between two groups \emph{a}
(\emph{initial}) and \emph{b} (\emph{post}). ``Zeros'' and ``ones'' signify
the absence and presence (respectively) of a sunspot group in a longitude or
latitude bin. The moderate variation in longitude is visible in the encoded
data (lines 11\,--\,17).  The \emph{post} link group is encoded with observations
nearest the limb ranked highest. Thus the ninth line of the data
representation encodes the longitude of group \emph{b} observed closest to
the east limb.  The line of ``ones'' to the right of centre indicates that this
group appears at a greater longitude than the \emph{initial} group. Similarly,
latitude indicated by colour is encoded in lines 27\,--\,33 for group \emph{a}
and lines 19\,--\,25 for group \emph{b}. Line 37 encodes the judgement on
whether or not the two groups shown are the same recurrent group; accordingly,
`zero' is false, `one' is true. Line 35 includes the Greenwich group numbers
of the two groups. This is not used in decision making, it is used to identify
groups after the neural network decision making process is complete.}

\label{fig:GPRtoNN}
\end{center}
\end{figure}

The encoding of initial and post groups takes place as follows. The
observation closest to the west limb is found for the initial group. For each
observation in time, sunspot groups are quantised into a given number of bins
for both latitude and longitude. Longitude and latitude are encoded
separately.  For the initial group, five latitude bins and seven longitude
bins are used. The first observation of the group is placed in the middle bin
(1) with the other bins at this time step set to zero (Figure
\ref{fig:GPRtoNN}, line 11 for longitude and line 27 for latitude).  Tracking
back from the west limb of the Sun, time steps are assumed to be 180/15
\textdeg/day (Figure \ref{fig:GPRtoNN}, lines 11\,--\,17 and lines 27\,--\,33). This
time step corresponds to the typical observed movement of a group due to solar
rotation. If no group is observed during a particular time step, all the bins
are set to zero (0).  Subsequent group observations are placed in bins
relative to the previous observation of the group. Bin widths are 1/2\textdeg\
and 1/7\textdeg\ for longitude and latitude respectively. This process is
repeated for seven time steps and is illustrated for group numbers 8929 and 8965
in Figure \ref{fig:GPRtoNN}.  Uncoded groups are visualised on the left, with
the corresponding network encoding on the right.

The process for treatment of the post group is similar.  For the post group, 9
latitude bins and 21 longitude bins are used (Figure \ref{fig:GPRtoNN}, lines
19\,--\,25 and lines 3\,--\,9). Again, seven observations in time are made. In
the case of the post group, however, all binning is performed relative to the
last observed longitude of the initial group. Extreme deviations are limited
at the bin furthest from the middle bin.

\subsection{Constructing Training Data}

In this study, the judgment of the first author (Henwood, 2008) was initially
used to decide if any two sunspot groups were close enough to represent
a recurrent group.  The dataset was balanced and a feed-forward neural
network system was then trained with these judgments. In this context, a
balanced dataset is one with an equal number of positive and negative
examples. The system generalises from the examples presented to it and
captures the uncertainty of the task.  When a previously unseen case of
possible recurrence is introduced, the trained neural network provides a
probability that this is indeed a case of recurrence.  

Solar cycle 15 (August 1913 -- August 1923) has been selected as the period to
identify recurrence manually, since this cycle is neither the longest nor the
shortest. It can also be described as a period when the overall sunspot group
number is neither particularly large nor small. In short, this cycle is chosen
because it is ``typical''.

Only sunspot groups in the \GPRu\ dataset are considered for recurrence.
Pairs of groups which appear to be linked by recurrence are selected for the
entire ten years of \GPRt. The training data, \GPRt, consists of 4073
examples, 621 of which are ``linked'' or $probability=1$, the remaining are
``unlinked'' or $probability=0$. Such a dataset, which contains unequal numbers
of true and false examples, is called ``unbalanced''. \inlinecite{provost2000}
highlights the problems associated with using unbalanced data with standard
machine learning algorithms. \inlinecite{fawcett2004} suggests the use of
receiver operating characteristic (ROC) curves to evaluate a classifier
trained on unbalanced data, while \inlinecite{lawrence1998} provide a number
of suggestions in order to balance unbalanced data. 

In this study, the training dataset is balanced using two techniques: firstly,
the addition of random noise (switching a single random adjacent bit in the
network representation) and secondly a technique based on \emph{a priori}
knowledge of the problem domain. This second technique uses the following
assumption: If one accepts that a given pair of groups are linked, any pair
which have a smaller unseen latitude and/or longitude deviation must also be
linked. Hence, new linked groups can be created from existing groups by
re-encoding a linked pair of groups with a reduced sensitivity than described
in Section \ref{sec:present}.  Using these two techniques, the training data
is balanced to produce 3756 positive examples and 3827 negative examples.

Ten-fold cross validation \cite{kohavi1995} has been used during training to
provide the learning and error rates. These are measured to identify network
designs that exhibit overfitting or underfitting \cite{moore2001}. Overfitting
is detected by observing that the error rate does not increase as learning is
taking place. Under-fitting is detected by observing that learning has
approached a stable minimum. Six network designs were constructed with varying
numbers of hidden layers and interconnects. These designs are documented in
the Master's Thesis of Henwood (2008), which also provides an assessment of
the performance of each network. The trained network that exhibits the least
overfitting or underfitting is presented with the \GPRu\ dataset and, after
classification, returns a new dataset with recurrent sunspots grouped and
labelled with the group number of the first observed sunspot group. This
dataset is called \GPRlp\ and contains 5374 group linkages.

A comparison of \GPRt\ with \GPRlp\ was then performed by hand.
This revealed 20 or so linked groups identified by the neural network that 
exhibited a large deviation of latitude or longitude during unseen passage.
We judged these to be physically unlikely but since there is no absolute
criterion for this classification these could in principle be identified as
linked groups by another observer. We have chosen a criterion with which to
filter the data.

A filter was constructed which removed linked groups that exhibited a
large latitude deviation ($>8.5$\textdeg) or the post link was not observed near
to the East limb, which corresponds to an large ($>19.5$\textdeg) longitude
deviation. This filter removed 17 linkages (3\%) from \GPRt\ and 244
(5\%) linkages from \GPRlp. This suggests that the majority of the
``physically unlikely'' groups were not a result of overfitting or
underfitting but are the result of subjectivity during selection of linked
groups for the training dataset \GPRt.

A final dataset, \GPRl\ was produced by applying the filter described above to
\GPRlp. The result is a dataset with contentious linkages removed. This
dataset is used throughout the remainder of this study and is available at the
UK Solar System Data Centre
(\url{http://www.ukssdc.ac.uk/wdcc1/greenwich/recurrence})

\section{Comparison with Manual Datasets}
\label{sec:comparison}

The UK Solar System Data Centre (\url{http://www.ukssdc.ac.uk}) maintains a complete
set of annals that were published by the Royal Greenwich Observatory; these
provided the source for the \GPR. As an appendix to the Greenwich
observations, a ``Catalogue of Recurrent Groups of Sun Spots, 1874 -- 1906'' was
compiled by A.S.D.~Maunder and published in 1909. Between 1916 and
1955, the ``Ledgers of Groups of Sunspots'' included two sections: ``Recurrent''
and ``Non-Recurrent''. Recurrent groups were also identified between 1907 and
1915 but tabulation of the information during that time was different.

While a digital version of these records is not known to exist, the method
used to compile recurrent spots is documented (in the
\emph{Greenwich Photoheliographic Results, 1956}):

\begin{quote}

Recurrent groups were selected upon the following plan, reference being made
to the General Catalogue:- If any spot when first seen was 60$^\circ$ or more
to the east of the central meridian, the catalogue and, if necessary, the
Daily Results also, were searched some fifteen to sixteen days earlier to
ascertain whether a spot group of similar heliographic longitude and latitude
was then near the west limb of the Sun. Similarly, if any spot group when
last seen was 60$^\circ$ or more to the west of the central meridian, a
search was made fifteen to sixteen days later. When there appeared to be a
case of probable continuity between groups in consecutive rotations of the
Sun, the character of the groups, their areas and their longitude and
latitude have been carefully compared before accepting them as recurrent
groups.

\end{quote}

Between 1874 and 1906, Maunder catalogued 624 recurring groups:  468 were seen
only in two rotations, 118 appeared in three rotations, 25 in four rotations,
12 in five rotations, and 1 (somewhat doubtfully) in six rotations.

The process of verification addresses the question of whether or not the work
described in this paper has been completed correctly. One approach is
to compare \GPRl\ with the recurrence observations that were published by the
\RGO.  Recurrence data for 1896 and 1958 were typed in and compared with
\GPRl. For interest, the performance of the neural network against the
training data, \GPRt, was also evaluated.

\begin{table}
\begin{tabular}{l r r r r }
\hline
Dataset & True Positives & True Negatives & False Positives & False Negatives \\
\hline
\GPRt    & 450 & 3799 & 34  & 170 \\
RGO 1958 & 50  & 2032 & 106 & 11  \\
RGO 1896 & 15  & 180  & 13  & 11  \\
\hline
\end{tabular}

\caption{Recurrent sunspot groups found by Henwood (2008), termed \GPRt, and
two other recurrent group datasets tabulated within the \RGO\ publications are
compared with \GPRl.  The columns of values, from left to right are: the
number of linked groups which the human and neural network classifier agree on
(True Positive count); the number of linkages which both the human and neural
network agree are not linked (True Negative count); the number of links which
the neural network classifies as true but the human does not (False Positive
count); the number of linkages the human thinks are true but the neural
network does not (False Negative count).}

\label{tab:verify}
\end{table}

Table \ref{tab:verify} compares these three different linked sunspot group
datasets with \GPRl\, using metrics for a confusion matrix \cite{kohavi1998}.
True positive counts are links which are in both the manual and \GPRl\
datasets.  False positives are links which are only in \GPRl.  False negatives
are links which are in the manual dataset but not in \GPRl.  True negatives
are defined as: $TN = Total-TP-FP-FN$. 

By rapidly constructing longitude-time diagrams that correspond to the
period of interest the instances of false classification have been
investigated individually. As may be expected, \GPRt\ data performs the best
overall. It does not achieve 100\% success compared to the human
classification. On inspection of the ``False Positive'' classifications made by
the neural network, all but one of these linkages were discovered to be
physically likely linkages which the human classifier had overlooked.  During
inspection by longitude--time diagrams it was observed that these ``False
Positive'' results occurred in dense complexes of many groups. The neural
network out-performs the human under these conditions.

The \RGO\ 1958 recurrence records show the highest number of false positives.
On investigating the set of examples classified as false positives, it was
observed that these were again a result of a complication arising during
observation of longitude--time diagrams. It should be noted that 1958 was a
year during which the sunspot number reached a record high. This level of
activity corresponds to a greater number of observations to display on a
longitude--time diagram, which increases the challenge to a human of identifying
all potentially linked groups. 

In addition, the method used during 1958 was restricted to providing only one
link between two given groups. However, the neural network classifier could
find multiple linkages between groups. Finally, the neural network classifier
could also make a link between groups even if one of the groups was only
visible for a single day. Such a linkage is apparently not permitted within
the 1958 recurrence dataset.

The RGO 1896 recurrence records showed the greatest discrepancy from the
neural network linked dataset. This is probably because the criteria used to
identify a recurrent group are different from both the 1958 and the \GPRt\
datasets. Maunder used heliographic position, allowing for a sunspot group to
remain unseen for some days and marking it as recurrent if another group was
observed at the same heliographic position. The method employed here is to
require a sunspot group to meet the ``unreliable observed'' criteria. Some of
the groups marked as recurrent by Maunder failed to be seen within 30$^\circ$
of the relevant limb. It should also be noted that 1958 was a year of
considerable activity on the Sun, whereas 1896 was relatively quiet, reducing
the opportunity to observe recurrence.

\section{Gnevyshev--Waldmeier Rule within Recurrent Groups}
\label{sec:gvrule}

After classifying recurrence, one might expect the \GPRlr\ dataset to exhibit some
of the well established physical characteristics of non-recurrent
sunspot groups. The Gnevyshev--Waldmeier rule
\cite{gnevyshev1938,waldmeier1955} states that sunspot group maximum area
$(A_0)$ and lifetime $(T)$ are proportional:

\begin{equation}
A_0 = \bar{D}_{GW}T\quad\quad \bar{D}_{GW} \approx 10 \mathrm{MSH day^{-1}}. \label{eqn:sizeage}
\end{equation}

\noindent where $\bar{D}_{GW}$ is the constant of proportionality measured in
millionths of the solar hemisphere (MSH) per day.

In their paper on sunspot decay, \inlinecite{petrovay1997} used the Debrecen
recurrence dataset \cite{dezso1987,dezso1997}. They choose to use groups which
``were born on the visible hemisphere and also died there'', so that their
lifetimes could be determined accurately. Since only one group satisfied this
criterion, they applied the Gnevyshev--Ringnes correction
to the remaining recurrent groups to make a
total dataset of 128 groups. The Gnevyshev--Ringnes correction provides the
probability that birth and death will occur in the visible hemisphere.  After
binning, they found a least squares linear fit of $\bar{D}_{GW} =$ 10.89
$\pm$0.48 MSH day$^{-1}$.

\begin{figure}
\begin{center}
\scalebox{1.0}{\input{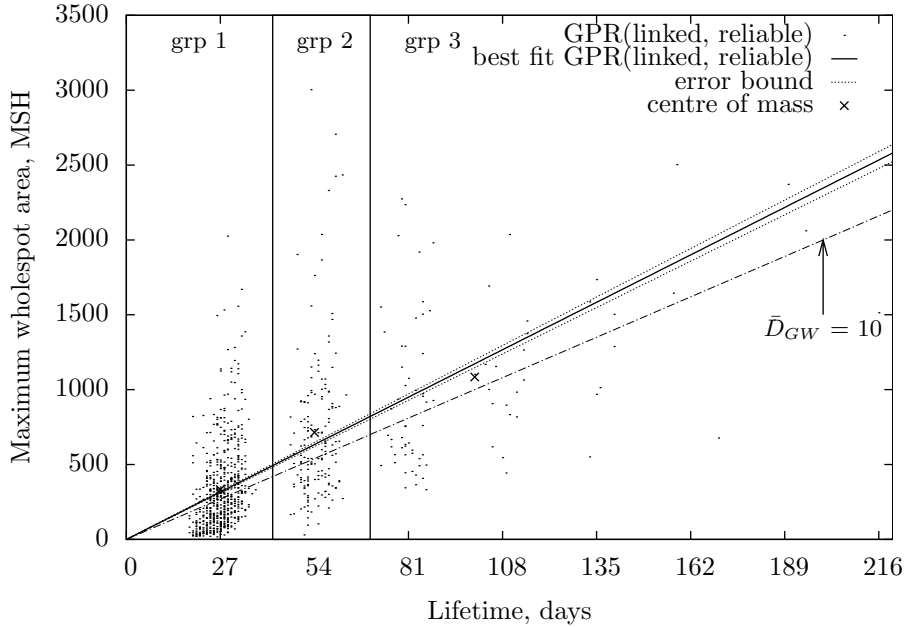}}

\caption{Recurrent sunspot group lifetime (measured in days) plotted
against maximum wholespot size measured in millionths of the solar hemisphere
(MSH). Lifetime is measured as the duration between first and last
observation of a sunspot group. The \GPRlr\ dataset contains 841 observations
and a linear fit of $\bar{D}_{GW}$ = 11.73 $\pm$0.26 is found. The data are
divided into three age categories (grp 1, grp 2 and grp 3) and the centre of
mass of each is indicated with a cross. The error bounds are marked by dashed
lines and $\bar{D}_{GW}$ = 10 is included for comparison.}

\label{fig:sizeage}
\end{center}
\end{figure}

There are 841 groups in \GPRlr\ that are reliably observed according to the
criteria that both the birth and death of a group must take place within
$\pm$60\textdeg\ of the solar central meridian. Using reliably observed
groups means the Gnevyshev--Ringnes correction is not required. Figure
\ref{fig:sizeage} shows a linear fit through \GPRlr\ of $\bar{D}_{GW}$ = 11.73
$\pm$0.26.

All of the lifetimes within \GPRlr\ are accurate to within one day. The
corresponding measurement of group maximum size is subject to some uncertainty
because rotation of the Sun carries each recurrent group out of sight for a
portion of its lifetime. The effect is that the value observed as the
maximum is either the true maximum or a smaller value. 

The data show a large scatter around the linear fit (Figure
\ref{fig:sizeage}). Regions between the bands of points (where no reliable
observation is possible because either the birth or death of the group is
unseen) make it somewhat difficult to determine the linear relationship. Three
age categories are defined as follows: grp 1, ages 17 to 45 days; grp 2, 46 to
72 days; grp 3, greater than 72 days.  The centre of mass of each of these
groups is indicated by a cross. In addition, since only recurrent sunspot
groups are examined, there are no data points between the origin and
$\approx$18 days. 

Compared to the data analysed by \inlinecite{petrovay1997}, the \GPRlr\ data
are more numerous. In addition, the \GPRlr\ data do not have a limit on the
maximum size measured and only recurrent groups are included in the fitting.
For these reasons, one should not expect exact agreement between
$\bar{D}_{GW}$ in Figure \ref{fig:sizeage} and values found from previous
investigations. In this study, $\bar{D}_{GW}$ has a smaller uncertainty but is
larger than previous estimates. The discrepancy between the value obtained in
this paper and the one found by \inlinecite{petrovay1997} is small, if
allowance is made for the appropriate error bounds, which suggests that for
recurrent groups $\bar{D}_{GW}$ is probably closer to 11 MSH day$^{-1}$ than 10
MSH day$^{-1}$.

\section{Temporal Variations of Sunspot Group Lifetime}
\label{sec:temporalvariations}

The \GPRl\ dataset presents a unique opportunity to investigate changes of
sunspot lifetime with time. Previous studies of this property were complicated 
for the reasons already outlined; namely, incomplete recurrence data compiled
by different individuals using different criteria.

\inlinecite{blanter2006} considered the topic of sunspot lifetime. They
performed a nonlinear study of the short-term correlation properties of solar
activity in order to reveal their long-lifetime variations.  Their method was
applied to \GPRu\ and allows for the problems associated with recurrence
within the dataset and short-lived sunspots. These authors mitigate such
factors by examining the population of sunspot lifetimes that are between 1
and 15 days. These observations were used to create a 22-year running
averaged series.

\begin{figure}
\begin{center}
\scalebox{1.0}{\input{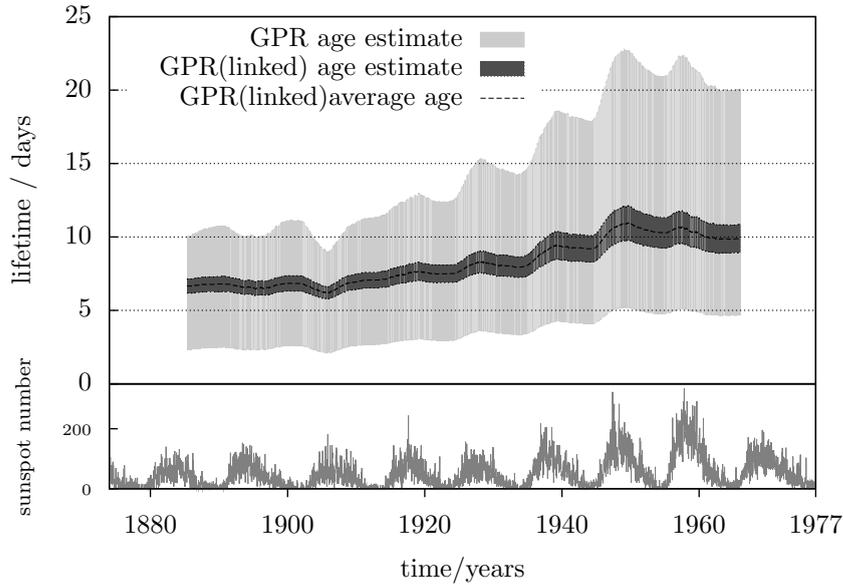}}

\caption{Lifetime of sunspot groups versus time, calculated for the
longest-lived sunspot on a given day for 22-year (8035 days) moving averages.
The \GPR\ age estimate takes no account of recurrent sunspot groups and is
plotted as a grey pattern contained within the bounds of error. \GPRl\ is
plotted as a dark grey solid region containing the bounds of uncertainty.
Error is calculated as the longest and shortest lifetime estimates of a
sunspot group lifetime.  Longest lifetime assumes that group birth was at the
earliest possible unseen time and death was at the latest possible unseen
time.  Shortest lifetime is calculated assuming the earliest observation in
\GPR\ is the birth and the latest observation in \GPR\ is the death.
Variations in sunspot number are shown to indicate individual solar cycles.}

\label{fig:ageVsTimeMean}
\end{center}
\end{figure}

One of the conclusions of the study by \inlinecite{blanter2006} was that they found
sunspot lifetime of \GPRu\ had increased over the duration of the
dataset. In addition, they were able to quantify the change in lifetime, which
increased by a factor of 1.4 over the interval from 1915 to 1940.
During this period, \GPRl\ indicates a change from below
to above average lifetimes, as shown in Figure \ref{fig:ageVsTimeMean}.

\inlinecite{blanter2006} discuss the problems associated with both small sunspot
groups, whose lifetime cannot be perfectly measured, and the lack of
observations from the invisible side of the Sun. Because of this, they
developed a technique that used sunspot group size and the
Gnevyshev--Waldmeier relationship to infer sunspot group lifetimes. Figure
\ref{fig:ageVsTimeMean} presents \GPR\ data (grey pattern region) and \GPRl\
(grey solid region) from our study.

Large errors are observed in the calculation of lifetime from \GPR\ data
containing recurrent groups. \GPRl\ alone contains much reduced error and a
trend can be observed. While \GPRl\ only contains groups that have a
sufficiently long lifetime to be observed on more than one solar rotation,
there are proportionally fewer such groups, which reduces the lifetime average
during the sample window.

\GPRl\ data shown in Figure \ref{fig:ageVsTimeMean} indicate a marked increase
between 1910 and 1950. \inlinecite{blanter2006} found that sunspot lifetime had
increased by a factor of 1.4 between 1915 and 1940. The results presented here
largely agree with that value. In addition, Figure \ref{fig:ageVsTimeMean}
suggests that the increase may extend over a longer interval and also augments
the work of \inlinecite{blanter2006} by introducing an uncertainty measure.

\section{Conclusions}

Neural networks have been applied previously to various problems in space
science. In particular, they have been used in the prediction of geomagnetic
phenomena
\cite{1992lundstedt,1994lundstedt,1995calvo,1996gleisner,1996williscroft,1996wu,1999weigel},
the classification of asteroid spectra \cite{1994howell}, and ionogram processing
\cite{1996galkin}. In addition, they have also been applied to the important
problem of automatically classifying sunspots from data obtained by processing
SOHO/MDI satellite images \cite{nguyen2006}. However, the authors are not
aware of the previous use of neural networks in studies of recurrent sunspot
groups.

It has been shown in this study that it is possible to train a neural network
to identify recurrent sunspot groups within the Greenwich Photoheliographic
Results (GPR), which extend over the long interval 1874 -- 1976 (Section
\ref{sec:method}).
Once trained, the neural network can often outperform a human classifier,
particularly when a large number of sunspot groups are present on the solar
disc at the same time. Since the neural network performs deterministically
when classifying a pair of linked sunspot groups, it operates with a
consistency of judgement that exceeds the sterling endeavours of various
individuals over more than a century.

Nevertheless, the neural network method has some limitations. These are most
clearly revealed when considering false-positive linkages (Section
\ref{sec:comparison}). In such cases, the ‘black-box’ nature of the decision
process becomes problematic and confidence in the neural network is partially
undermined. In the method discussed in this paper, post-processing of the data
is used to reduce false-positive linkages but in any future study it might be
valuable to investigate alternative machine learning systems.

Despite these limitations, the particular choice of network and data reduction
procedure employed in this study could also be applied to Rome Daily Sunspot
Reports (1958\,--\,2000), USSR Station Data (1968\,--\,1991), Mount Wilson Individual
Sunspot Data (1917\,--\,1985), Kodaikanal Individual Sunspot Data (1906\,--\,1987) and
Greenwich/Debrecen Observations (1874\,--\,2007), \emph{without re-training the
existing neural network}. All of these datasets are available through either
the National Geophysical Data Centre (\url{http://www.ngdc.noaa.gov}) or the UK
Solar System Data Centre (\url{http://www.ukssdc.ac.uk}).

The constants of proportionality in the Gnevyshev--Waldmeier rule ($A_0 =
\bar{D}_{GW}T$) derived in this study (Section \ref{sec:gvrule}) have been
found to be greater than previous estimates. \inlinecite{petrovay1997}, who
used information on recurrent sunspot groups extracted from the Debrecen
Photoheliographic Results, also found a value of $\bar{D}_{GW}$ that was
larger (10.89 MSH day$^{-1}$) than previous estimates.  In a study of sunspot
group lifetimes, \inlinecite{zuccarello1993} presented results which show a
change in the rotation rate between short-lived and long-lived (11-day old)
sunspot groups, pointing to a difference in ``aggregation capability'' of the
group within the convection zone. The present results suggest that there may
be some additional physics present within longer lived groups in the \GPRlr\
dataset and thereby imply that this matter warrants further investigation.

Evidence has been found for an increase in the lifetime of recurrent
sunspot groups by a factor of about 1.4 between 1915 and 1940 (Section
\ref{sec:temporalvariations}),
which is in excellent agreement with the result obtained by
\inlinecite{blanter2006}. Indeed, this increase in lifetime actually occurs over
a longer period (1915 -- 1950) than previously thought and there is also
provisional evidence for a slight decrease in lifetime between 1950 and 1965
(see Figure \ref{fig:ageVsTimeMean}).

Solar changes over periods longer than a few decades are currently of
considerable interest as the solar output is a significant input to climate
models \cite{ssac2005}. The Gleissberg cycle is detected in sunspot number and
has a measured period of approximately 80-120 years
\cite{1967gleissberg,hoyt1994,garcia1998}.  \inlinecite{garcia1998} estimated
the most recent Gleissberg cycle minima to be around 1900 and the maxima
around 1965. The results presented here suggest that, if the lifetimes of
recurrent sunspot groups are a good proxy for the Gleissberg cycle, the maxima
occurred some years before 1965 during the 1950s.  Further study of this topic would be
facilitated by applying the trained neural network to sunspot data in the
interval 1977\,--\,2009.

The analysis of historical sunspot observations using automated methods, such
as the neural network method presented in this paper, is required to estimate
the (true) total number of sunspots emerging on the solar surface during the
solar cycle. Numerical simulations of the centennial evolution of magnetic
flux on the solar surface scale the sunspot emergence rate to the total
sunspot number, measured without considering recurrent sunspots and their
variable lifetimes in great detail \cite{wang2002}. A sunspot number that
corrects for the recurrence of long-lived sunspots may improve the numerical
predictions by providing a better description of emergence. The numerical
simulations over centennial scales could be compared to the most recent
estimates of long-term variation of the open magnetic flux on the solar
surface \cite{rouillard2007}. Neural networks could also be optimised to
detect the location of the umbra and penumbra of sunspots and to estimate the
total magnetic flux inside these large-scale active regions. Such a rough
estimate of the magnetic flux on the photosphere could then be used to
estimate the magnetic topology of the corona using simplified potential field
source surface models.

\section*{Acknowledgements}

The authors would like to thank one anonymous reviewer for their valuable
suggestions and comments.  R. Henwood thanks Matthew Wild, Sarah James,
Richard Stamper, and Alexis Rouillard for their support and helpful criticism.
Part of the research reported in this paper was undertaken by RH in partial
fulfilment of the requirements for the MSc Degree at the University of
Warwick. While studying for this degree, he was supported by the Space Science
and Technology Department of the Rutherford Appleton Laboratory.  S. Fell,
then at a local college, spent a month working with RH on evaluating machine
learning software, compiling training data, and designing networks.

\bibliographystyle{./spr-mp-sola-cnd}
\bibliography{SOLA_1023.bbl}

\begin{thebibliography}{43}
\ifx \bisbn   \undefined \def \bisbn  #1{ISBN #1}   \fi
\ifx \binits  \undefined \def \binits#1{#1} \fi
\ifx \bauthor  \undefined \def \bauthor#1{#1} \fi
\ifx \batitle  \undefined \def \batitle#1{#1} \fi
\ifx \bjtitle  \undefined \def \bjtitle#1{#1} \fi
\ifx \bvolume  \undefined \def \bvolume#1{#1} \fi
\ifx \byear  \undefined \def \byear#1{#1} \fi
\ifx \bissue  \undefined \def \bissue#1{#1} \fi
\ifx \bfpage  \undefined \def \bfpage#1{#1} \fi
\ifx \blpage  \undefined \def \blpage #1{#1} \fi
\ifx \burl  \undefined \def \burl#1{#1} \fi
\ifx \binterref  \undefined \def \binterref#1{#1} \fi
\ifx \betal  \undefined \def \betal#1{#1} \fi
\ifx \binstitute  \undefined \def \binstitute#1{#1} \fi
\ifx \bctitle  \undefined \def \bctitle#1{#1} \fi
\ifx \beditor  \undefined \def \beditor#1{#1} \fi
\ifx \bpublisher  \undefined \def \bpublisher#1{#1} \fi
\ifx \bbtitle  \undefined \def \bbtitle#1{#1} \fi
\ifx \bedition  \undefined \def \bedition#1{#1} \fi
\ifx \bseriesno  \undefined \def \bseriesno#1{#1} \fi
\ifx \blocation  \undefined \def \blocation#1{#1} \fi
\ifx \bsertitle  \undefined \def \bsertitle#1{#1} \fi
\ifx \bsnm \undefined \def \bsnm#1{#1} \fi
\ifx \bsuffix \undefined \def \bsuffix#1{#1} \fi
\ifx \bparticle \undefined \def \bparticle#1{#1} \fi
\ifx \barticle \undefined \def \barticle#1{#1} \fi
\ifx \botherref \undefined \def \botherref #1{#1} \fi
\ifx \url \undefined \def \url#1{\textsf{#1}} \fi
\ifx \bchapter \undefined \def \bchapter#1{#1} \fi
\ifx \bbook \undefined \def \bbook#1{#1} \fi
\ifx \bcomment \undefined \def \bcomment#1{#1} \fi
\ifx \oauthor \undefined \def \oauthor#1{#1} \fi
\ifx \citeauthoryear \undefined \def \citeauthoryear#1{#1} \fi
\def \endbibitem {}

\bibitem[\protect\citeauthoryear{{Balmaceda}, {Krivova}, and
  {Solanki}}{2007}]{Balmaceda2007}
\begin{barticle}
\bauthor{\bsnm{{Balmaceda}},~\binits{L.}},
  \bauthor{\bsnm{{Krivova}},~\binits{N.A.}},
  \bauthor{\bsnm{{Solanki}},~\binits{S.K.}}:
\byear{2007}, \textit{\bjtitle{Adv.~Space Res.}}
  \textbf{\bvolume{40}}(\bissue{7}), \bfpage{986 }.
\end{barticle}
\endbibitem

\bibitem[\protect\citeauthoryear{{Balthasar}, {V{\'a}zquez}, and
  {W{\"o}hl}}{1986}]{balthasar1986}
\begin{barticle}
\bauthor{\bsnm{{Balthasar}},~\binits{H.}},
  \bauthor{\bsnm{{V{\'a}zquez}},~\binits{M.}},
  \bauthor{\bsnm{{W{\"o}hl}},~\binits{H.}}:
\byear{1986}, \textit{\bjtitle{Astron.~Astrophys.}} \textbf{\bvolume{155}},
  \bfpage{87}.
\end{barticle}
\endbibitem

\bibitem[\protect\citeauthoryear{{Becker}}{1955}]{becker1955}
\begin{barticle}
\bauthor{\bsnm{{Becker}},~\binits{U.}}:
\byear{1955}, \textit{\bjtitle{Z.~Astrophys.}} \textbf{\bvolume{37}},
  \bfpage{47}.
\end{barticle}
\endbibitem

\bibitem[\protect\citeauthoryear{{{Blanter}} {\it et~al.}}{2006}]{blanter2006}
\begin{barticle}
\bauthor{\bsnm{{Blanter}},~\binits{E.M.}}, \bauthor{\bsnm{{Le
  Mou{\"e}l}},~\binits{J.-L.}}, \bauthor{\bsnm{{Perrier}},~\binits{F.}},
  \bauthor{\bsnm{{Shnirman}},~\binits{M.G.}}:
\byear{2006}, \textit{\bjtitle{Solar~Phys.}} \textbf{\bvolume{237}},
  \bfpage{329}.
\end{barticle}
\endbibitem

\bibitem[\protect\citeauthoryear{{Calvo}, {Ceccato}, and
  {Piacentini}}{1995}]{1995calvo}
\begin{barticle}
\bauthor{\bsnm{{Calvo}},~\binits{R.A.}},
  \bauthor{\bsnm{{Ceccato}},~\binits{H.A.}},
  \bauthor{\bsnm{{Piacentini}},~\binits{R.D.}}:
\byear{1995}, \textit{\bjtitle{Astrophys.~J.}} \textbf{\bvolume{444}},
  \bfpage{916}.
\end{barticle}
\endbibitem

\bibitem[\protect\citeauthoryear{{Castenmiller}, {Zwaan}, and {van der
  Zalm}}{1986}]{castenmiller1986}
\begin{barticle}
\bauthor{\bsnm{{Castenmiller}},~\binits{M.J.M.}},
  \bauthor{\bsnm{{Zwaan}},~\binits{C.}}, \bauthor{\bsnm{{van der
  Zalm}},~\binits{E.B.J.}}:
\byear{1986}, \textit{\bjtitle{Solar~Phys.}} \textbf{\bvolume{105}},
  \bfpage{237}.
\end{barticle}
\endbibitem

\bibitem[\protect\citeauthoryear{{Dezs{\H{o}}}, {Gerlei}, and
  {Kov{\'a}cs}}{1987}]{dezso1987}
\begin{botherref}
\oauthor{\bsnm{{Dezs{\H{o}}}},~\binits{L.}},
  \oauthor{\bsnm{{Gerlei}},~\binits{O.}},
  \oauthor{\bsnm{{Kov{\'a}cs}},~\binits{{\'A}.}}:
1987, {\emph{Debrecen Photoheliographic Results for the Year 1977}}, Heliogr. Series No.
  1. Publ. Debrecen Obs., Debrecen.
\end{botherref}
\endbibitem

\bibitem[\protect\citeauthoryear{{Dezs{\H{o}}}, {Gerlei}, and
  {Kov{\'a}cs}}{1997}]{dezso1997}
\begin{botherref}
\oauthor{\bsnm{{Dezs{\H{o}}}},~\binits{L.}},
  \oauthor{\bsnm{{Gerlei}},~\binits{O.}},
  \oauthor{\bsnm{{Kov{\'a}cs}},~\binits{{\'A}.}}:
1997, {\emph{Debrecen Photoheliographic Results for the Year 1978}}, Heliogr. Series No.
  2. Publ. Debrecen Obs., Debrecen. \textsf{\url{ftp://fenyi.sci.klte.hu/pub/DPR/1978/}}.
\end{botherref}
\endbibitem

\bibitem[\protect\citeauthoryear{{Fawcett}}{2004}]{fawcett2004}
\begin{botherref}
\oauthor{\bsnm{{Fawcett}},~\binits{T.}}:
2004, {ROC Graphs: Notes and Practical Considerations for Researchers}.
Technical Report, HP Laboratories, Palo Alto.
\end{botherref}
\endbibitem

\bibitem[\protect\citeauthoryear{{Fligge} and {Solanki}}{1997}]{fligge1997}
\begin{barticle}
\bauthor{\bsnm{{Fligge}},~\binits{M.}},
  \bauthor{\bsnm{{Solanki}},~\binits{S.K.}}:
\byear{1997}, \textit{\bjtitle{Solar~Phys.}} \textbf{\bvolume{173}},
  \bfpage{427}.
\end{barticle}
\endbibitem

\bibitem[\protect\citeauthoryear{{Friis-Christensen} and
  {Svensmark}}{1997}]{svensmark1997}
\begin{barticle}
\bauthor{\bsnm{{Friis-Christensen}},~\binits{E.}},
  \bauthor{\bsnm{{Svensmark}},~\binits{H.}}:
\byear{1997}, \textit{\bjtitle{Adv.~Space Res.}} \textbf{\bvolume{20}},
  \bfpage{913}.
\end{barticle}
\endbibitem

\bibitem[\protect\citeauthoryear{{{Galkin}} {\it et~al.}}{1996}]{1996galkin}
\begin{barticle}
\bauthor{\bsnm{{Galkin}},~\binits{I.A.}},
  \bauthor{\bsnm{{Reinisch}},~\binits{B.W.}},
  \bauthor{\bsnm{{Ososkov}},~\binits{G.A.}},
  \bauthor{\bsnm{{Zaznobina}},~\binits{E.G.}},
  \bauthor{\bsnm{{Neshyba}},~\binits{S.P.}}:
\byear{1996}, \textit{\bjtitle{Radio Science}} \textbf{\bvolume{31}},
  \bfpage{1119}.
\end{barticle}
\endbibitem

\bibitem[\protect\citeauthoryear{{Garcia} and {Mouradian}}{1998}]{garcia1998}
\begin{barticle}
\bauthor{\bsnm{{Garcia}},~\binits{A.}},
\bauthor{\bsnm{{Mouradian}},~\binits{Z.}}:
\byear{1998}, \textit{\bjtitle{Solar~Phys.}} \textbf{\bvolume{180}},
\bfpage{495}.
\end{barticle}
\endbibitem

\bibitem[\protect\citeauthoryear{{Gleisner}, {Lundstedt}, and
  {Wintoft}}{1996}]{1996gleisner}
\begin{barticle}
\bauthor{\bsnm{{Gleisner}},~\binits{H.}},
  \bauthor{\bsnm{{Lundstedt}},~\binits{H.}},
  \bauthor{\bsnm{{Wintoft}},~\binits{P.}}:
\byear{1996}, \textit{\bjtitle{Annales Geophysicae}} \textbf{\bvolume{14}},
  \bfpage{679}.
\end{barticle}
\endbibitem

\bibitem[\protect\citeauthoryear{{Gleissberg}}{1967}]{1967gleissberg}
\begin{barticle}
\bauthor{\bsnm{{Gleissberg}},~\binits{W.}}:
\byear{1967}, \textit{\bjtitle{Solar~Phys.}} \textbf{\bvolume{2}}, \bfpage{231}.
\end{barticle}
\endbibitem


\bibitem[\protect\citeauthoryear{{Gnevyshev}}{1938}]{gnevyshev1938}
\begin{barticle}
\bauthor{\bsnm{{Gnevyshev}},~\binits{M.N.}}:
\byear{1938}, \textit{\bjtitle{Izvestiya Glavnoj Astronomicheskoj Observatorii
  v Pulkove}} \textbf{\bvolume{16}}, \bfpage{36}.
\end{barticle}
\endbibitem

\bibitem[\protect\citeauthoryear{{{Haigh}} {\it et~al.}}{2005}]{ssac2005}
\begin{botherref}
\oauthor{\bsnm{{Haigh}}~\binits{J.D.}},
\oauthor{\bsnm{{Lockwood}}~\binits{M.}},
\oauthor{\bsnm{{Giampapa}}~\binits{M.S.}},
\oauthor{\bsnm{{R{\"u}edi}}~\binits{I.}},
\oauthor{\bsnm{{G{\"u}del}}~\binits{M.}},
\oauthor{\bsnm{{Schmutz}}~\binits{W.}}:
2005, {The Sun, Solar Analogs and the Climate}. Springer, Berlin.
\end{botherref}
\endbibitem


\bibitem[\protect\citeauthoryear{{Henwood}}{2008}]{henwood2008}
\begin{botherref}
\oauthor{\bsnm{{Henwood}},~\binits{R.}}:
2008, {\emph{An Investigation into Recurrence in Greenwich Photoheliographic Results
  1874--1976}}.
Master's Thesis, Centre for Fusion, Space and Astrophysics, Univ. Warwick.
\end{botherref}
\endbibitem

\bibitem[\protect\citeauthoryear{{Howell}, {Mer\'{e}nyi}, and
  {Lebofsky}}{1994}]{1994howell}
\begin{barticle}
\bauthor{\bsnm{{Howell}},~\binits{E.S.}},
  \bauthor{\bsnm{{Mer\'{e}nyi}},~\binits{E.}},
  \bauthor{\bsnm{{Lebofsky}},~\binits{L.A.}}:
\byear{1994}, \textit{\bjtitle{J.~Geophys.~Res.}} \textbf{\bvolume{99}},
  \bfpage{10847}.
\end{barticle}
\endbibitem

\bibitem[\protect\citeauthoryear{{Hoyt}, {Schatten}, and
{Nesmes-Ribes}}{1994}]{hoyt1994}
\begin{barticle}
\bauthor{\bsnm{{Hoyt}},~\binits{D.V.}},
\bauthor{\bsnm{{Schatten}},~\binits{K.H.}},
\bauthor{\bsnm{{Nesmes-Ribes}},~\binits{E.}}:
\byear{1994}, \textit{\bjtitle{Geophys.~Res.~Lett.}} \textbf{\bvolume{21}},
\bfpage{2067}.
\end{barticle}
\endbibitem


\bibitem[\protect\citeauthoryear{Kohavi}{1995}]{kohavi1995}
\begin{bchapter}
\bauthor{\bsnm{Kohavi},~\binits{R.}}:
\byear{1995}, In: \bbtitle{\textit{Proceedings of the Fourteenth International
  Joint Conferences on Artificial Intelligence}, {\bf{Vol. 2}}}, \bpublisher{Morgan
  Kaufmann}, \blocation{Montreal}, \bfpage{1137}.
\end{bchapter}
\endbibitem

\bibitem[\protect\citeauthoryear{Kohavi and Provost}{1998}]{kohavi1998}
\begin{barticle}
\bauthor{\bsnm{Kohavi},~\binits{R.}}, \bauthor{\bsnm{Provost},~\binits{F.}}:
\byear{1998}, \textit{\bjtitle{Mach.~Learn.}}
  \textbf{\bvolume{30}}(\bissue{2/3}), \bfpage{271}. \bcomment{(Editorial for
  the Special Issue on Applications of Machine Learning and the Knowledge
  Discovery Process)}.
\end{barticle}
\endbibitem

\bibitem[\protect\citeauthoryear{{Kopeck{\'y}}}{1985}]{kopecky1985}
\begin{barticle}
\bauthor{\bsnm{{Kopeck{\'y}}},~\binits{M.}}:
\byear{1985}, \textit{\bjtitle{Bull. Astronom. Inst. Czech.}} \textbf{\bvolume{36}}, \bfpage{359}.
\end{barticle}
\endbibitem

\bibitem[\protect\citeauthoryear{{{Lawrence}} {\it
  et~al.}}{1998}]{lawrence1998}
\begin{bchapter}
\bauthor{\bsnm{{Lawrence}},~\binits{S.}},
  \bauthor{\bsnm{{Burns}},~\binits{I.}},
  \bauthor{\bsnm{{Back}},~\binits{A.D.}},
  \bauthor{\bsnm{{Tsoi}},~\binits{A.C.}},
  \bauthor{\bsnm{{Giles}},~\binits{C.L.}}:
\byear{1998}, In: \textit{\bbtitle{{Neural Networks: Tricks of the Trade}}}.
  \bpublisher{Springer}, \blocation{London, UK}, \bfpage{299}.
\end{bchapter}
\endbibitem

\bibitem[\protect\citeauthoryear{{Lean}, {Beer}, and
  {Bradley}}{1995}]{lean1995}
\begin{barticle}
\bauthor{\bsnm{{Lean}},~\binits{J.}}, \bauthor{\bsnm{{Beer}},~\binits{J.}},
  \bauthor{\bsnm{{Bradley}},~\binits{R.}}:
\byear{1995}, \textit{\bjtitle{Geophys.~Res.~Lett.}} \textbf{\bvolume{22}},
  \bfpage{3195}.
\end{barticle}
\endbibitem

\bibitem[\protect\citeauthoryear{{{Luhmann}} {\it et~al.}}{2002}]{luhmann2002}
\begin{barticle}
\bauthor{\bsnm{{Luhmann}},~\binits{J.G.}}, \bauthor{\bsnm{{Li}},~\binits{Y.}},
  \bauthor{\bsnm{{Arge}},~\binits{C.N.}},
  \bauthor{\bsnm{{Gazis}},~\binits{P.R.}},
  \bauthor{\bsnm{{Ulrich}},~\binits{R.}}:
\byear{2002}, \textit{\bjtitle{J.~Geophys.~Res.}} \textbf{\bvolume{107}},
  \bfpage{1154}.
\end{barticle}
\endbibitem

\bibitem[\protect\citeauthoryear{{Lundstedt}}{1992}]{1992lundstedt}
\begin{barticle}
\bauthor{\bsnm{{Lundstedt}},~\binits{H.}}:
\byear{1992}, \textit{\bjtitle{Planet.~Space~Sci.}} \textbf{\bvolume{40}},
  \bfpage{457}.
\end{barticle}
\endbibitem

\bibitem[\protect\citeauthoryear{{Lundstedt} and
  {Wintoft}}{1994}]{1994lundstedt}
\begin{barticle}
\bauthor{\bsnm{{Lundstedt}},~\binits{H.}},
  \bauthor{\bsnm{{Wintoft}},~\binits{P.}}:
\byear{1994}, \textit{\bjtitle{Ann. Geophys.}} \textbf{\bvolume{12}},
  \bfpage{19}.
\end{barticle}
\endbibitem

\bibitem[\protect\citeauthoryear{{Meeus}}{1991}]{1991meeus}
\begin{botherref}
\oauthor{\bsnm{{Meeus}},~\binits{J.H.}}:
1991, \textit{Astronomical Algorithms}. Willmann-Bell, Richmond, VA
\end{botherref}
\endbibitem

\bibitem[\protect\citeauthoryear{{Moore}}{2001}]{moore2001}
\begin{botherref}
\oauthor{\bsnm{{Moore}},~\binits{A.W.}}:
2001, Cross-validation for detecting and preventing overfitting.
  \textsf{\url{http://www.cs.cmu.edu/afs/cs/user/awm/web/tutorials/overfit10.p%
df}}.
\end{botherref}
\endbibitem

\bibitem[\protect\citeauthoryear{{Nguyen} {\it et~al.}}{2006}]{nguyen2006}
\begin{barticle}
\bauthor{\bsnm{Nguyen},~\binits{T.T.}}, \bauthor{\bsnm{Willis},~\binits{C.P.}},
  \bauthor{\bsnm{Paddon},~\binits{D.J.}},
  \bauthor{\bsnm{Nguyen},~\binits{S.H.}},
  \bauthor{\bsnm{Nguyen},~\binits{H.S.}}:
\byear{2006}, \textit{\bjtitle{Fundam. Inform.}}
  \textbf{\bvolume{72}}(\bissue{1-3}), \bfpage{295}.
\end{barticle}
\endbibitem

\bibitem[\protect\citeauthoryear{{Petrovay} and {van
  Driel-Gesztelyi}}{1997}]{petrovay1997}
\begin{barticle}
\bauthor{\bsnm{{Petrovay}},~\binits{K.}}, \bauthor{\bsnm{{van
  Driel-Gesztelyi}},~\binits{L.}}:
\byear{1997}, \textit{\bjtitle{Solar~Phys.}} \textbf{\bvolume{176}},
  \bfpage{249}.
\end{barticle}
\endbibitem

\bibitem[\protect\citeauthoryear{{Pierce} and {Slaughter}}{1977}]{Pierce1977}
\begin{barticle}
\bauthor{\bsnm{{Pierce}},~\binits{A.K.}},
  \bauthor{\bsnm{{Slaughter}},~\binits{C.D.}}:
\byear{1977}, \textit{\bjtitle{Solar~Phys.}} \textbf{\bvolume{51}}, \bfpage{25}.
\end{barticle}
\endbibitem

\bibitem[\protect\citeauthoryear{Provost}{2000}]{provost2000}
\begin{bbook}
\bauthor{\bsnm{Provost},~\binits{F.}}:
\byear{2000}, \textit{\bbtitle{{Machine Learning from Imbalanced Data Sets 101
  (Extended Abstract)}}}, \bpublisher{{Association for the Advancement of
  Artificial Intelligence Workshop on Imbalanced Data Sets}},
  \blocation{Austin, Texas}.
\end{bbook}
\endbibitem

\bibitem[\protect\citeauthoryear{{Ringnes}}{1964}]{ringnes1964}
\begin{barticle}
\bauthor{\bsnm{{Ringnes}},~\binits{T.S.}}:
\byear{1964}, \textit{\bjtitle{Astrophys.~Norv.}} \textbf{\bvolume{9}},
  \bfpage{95}.
\end{barticle}
\endbibitem

\bibitem[\protect\citeauthoryear{{Rouillard}, {Lockwood}, and
  {Finch}}{2007}]{rouillard2007}
\begin{barticle}
\bauthor{\bsnm{{Rouillard}},~\binits{A.P.}},
  \bauthor{\bsnm{{Lockwood}},~\binits{M.}},
  \bauthor{\bsnm{{Finch}},~\binits{I.}}:
\byear{2007}, \textit{\bjtitle{J.~Geophys.~Res.}}
  \textbf{\bvolume{112}}, \bfpage{A5103}, doi:10.1029/2006JA012130.
\end{barticle}
\endbibitem

\bibitem[\protect\citeauthoryear{Royal Greenwich Observatory}{1980}]{rgo1980}
\begin{bbook}
\bauthor{\bsnm{Royal Greenwich Observatory}}:
\byear{1980}, \bbtitle{\textit{Royal Observatory Annals, Photoheliographic Results 1972-1976}}, 
    \bpublisher{Royal Greenwich Observatory}, \blocation{Herstmonceux}.
\end{bbook}
\endbibitem


\bibitem[\protect\citeauthoryear{{Schr{\"{o}}ter}}{1985}]{schroeter1985}
\begin{barticle}
\bauthor{\bsnm{{Schr{\"{o}}ter}},~\binits{E.H.}}:
\byear{1985}, \textit{\bjtitle{Solar~Phys.}} \textbf{\bvolume{100}},
  \bfpage{141}.
\end{barticle}
\endbibitem

\bibitem[\protect\citeauthoryear{{Solanki}}{2003}]{solanki2003}
\begin{barticle}
\bauthor{\bsnm{{Solanki}},~\binits{S.K.}}:
\byear{2003}, \textit{\bjtitle{Astron.~Astrophys.~Rev.}} \textbf{\bvolume{11}},
  \bfpage{153}.
\end{barticle}
\endbibitem

\bibitem[\protect\citeauthoryear{{Thompson} {\emph{et~al.}}}{1996}]{thompson1996}
\begin{barticle}
\bauthor{\bsnm{Thompson},~\binits{M.J.}}, \bauthor{\bsnm{Toomre},~\binits{J.}},
  \bauthor{\bsnm{Anderson},~\binits{E.R.}},
  \bauthor{\bparticle{et~}\bsnm{al.}}:
\byear{1996}, \textit{\bjtitle{Science}} \textbf{\bvolume{272}}(\bissue{5266}),
  \bfpage{1300}.
\end{barticle}
\endbibitem

\bibitem[\protect\citeauthoryear{{Waldmeier}}{1955}]{waldmeier1955}
\begin{botherref}
\oauthor{\bsnm{{Waldmeier}},~\binits{M.}}:
1955, \textit{Ergebnisse und Probleme der Sonnenforschung.} Geest and Portig, Leipzig.
\end{botherref}
\endbibitem

\bibitem[\protect\citeauthoryear{{Wang}, {Lean}, and
  {Sheeley}}{2002}]{wang2002}
\begin{barticle}
\bauthor{\bsnm{{Wang}},~\binits{Y.-M.}}, \bauthor{\bsnm{{Lean}},~\binits{J.}},
  \bauthor{\bsnm{{Sheeley}},~\binits{N.R.}}:
\byear{2002}, \textit{\bjtitle{Astrophys.~J.~Lett.}} \textbf{\bvolume{577}},
  \bfpage{L53}.
\end{barticle}
\endbibitem

\bibitem[\protect\citeauthoryear{{{Weigel}} {\it et~al.}}{1999}]{1999weigel}
\begin{barticle}
\bauthor{\bsnm{{Weigel}},~\binits{R.S.}},
  \bauthor{\bsnm{{Horton}},~\binits{W.}},
  \bauthor{\bsnm{{Tajima}},~\binits{T.}},
  \bauthor{\bsnm{{Detman}},~\binits{T.}}:
\byear{1999}, \textit{\bjtitle{Geophys.~Res.~Lett.}} \textbf{\bvolume{26}},
  \bfpage{1353}.
\end{barticle}
\endbibitem

\bibitem[\protect\citeauthoryear{Williscroft and Poole}{1996}]{1996williscroft}
\begin{barticle}
\bauthor{\bsnm{Williscroft},~\binits{L.-A.}},
  \bauthor{\bsnm{Poole},~\binits{A.W.V.}}:
\byear{1996}, \textit{\bjtitle{Geophys.~Res.~Lett.}} \textbf{\bvolume{23}},
  \bfpage{3659}.
\end{barticle}
\endbibitem

\bibitem[\protect\citeauthoryear{{Wu} and {Lundstedt}}{1996}]{1996wu}
\begin{barticle}
\bauthor{\bsnm{{Wu}},~\binits{J.-G.}},
  \bauthor{\bsnm{{Lundstedt}},~\binits{H.}}:
\byear{1996}, \textit{\bjtitle{Geophys.~Res.~Lett.}} \textbf{\bvolume{23}},
  \bfpage{319}.
\end{barticle}
\endbibitem

\bibitem[\protect\citeauthoryear{{Zuccarello}}{1993}]{zuccarello1993}
\begin{barticle}
\bauthor{\bsnm{{Zuccarello}},~\binits{F.}}:
\byear{1993}, \textit{\bjtitle{Astron.~Astrophys.}} \textbf{\bvolume{272}},
  \bfpage{587}.
\end{barticle}
\endbibitem









\end{thebibliography}

\end{article}
\end{document}